\documentclass[twocolumn,showpacs,setspace,preprintnumbers,amsmath,amssymb]{revtex4}
\usepackage{graphicx}
\usepackage{dcolumn}
\usepackage{bm}
\usepackage{hyperref} 
\usepackage{latexsym}

\begin{document}
\title{Numerical Simulation of Particle Flow in a Sand Trap}
\author{A. D. Ara\'ujo$^{1}$, J. S. Andrade Jr.$^{1,3}$, L. P. Maia$^{2}$ and H. J. Herrmann$^{1,3}$}
\address {$1$ Departamento de F\'{\i}sica, Universidade Federal do
Cear\'a, 60451-970 Fortaleza, Cear\'a, Brazil.\\}
\address {$2$ Instituto de Ci\^encias do Mar (LABOMAR), 
Universidade Federal do Cear\'a, 60165-081 Fortaleza, Cear\'a, Brazil.\\}
\address {$3$ Computational Physics, IfB, ETH-Hoenggerberg, Schafmattstrasse 6, 8093 Z$\ddot{u}$rich, Switzerland.}
\date{\today}

\begin{abstract}

Sand traps are used to measure Aeolian flux. Since they
modify the surrounding wind velocity field their gauging
represents an important challenge. We use numerical
simulations under the assumption of homogeneous turbulence
based on FLUENT to systematically study the flow field and
trapping efficiency of one of the most common devices based
on a hollow cylinder with two slits. In particular, we
investigate the dependence on the wind speed, the Stokes
number, the permeability of the membrane on the slit and the
saltation height.
 
\end{abstract} 
\pacs{45.70.Mg,47.55.Kf,47.27.-i,83.80.Hj}
\maketitle

\section{Introduction}

Dune motion, sand encroachment and desertification are based
on sand transport by wind. Saltation is the basic mechanism
of Aeolian sand flux~\cite{Bagnold,Owen,Roux}. The simplest
and most common devices to measure it in the field are so
called sand traps. They consist of cavities into which the
sand bearing wind enters, drops the sand inside and then
leaves again without the sand. In that way they accumulate
inside the sand that would have crossed them during a given
time, giving a measure for the flux through their cross
section. The difficulty to quantitative gauge them arises in
properly estimating this cross section because the trap
being an extended fixed object modifies considerably the
wind velocity field in its surrounding. The issue boils down to
understand how much wind actually enters into the cavity and
therefore strongly depends on the geometrical shape of the
trap.

\begin{figure}
\begin{center}
\includegraphics[width=9.0cm,height=10.0cm]{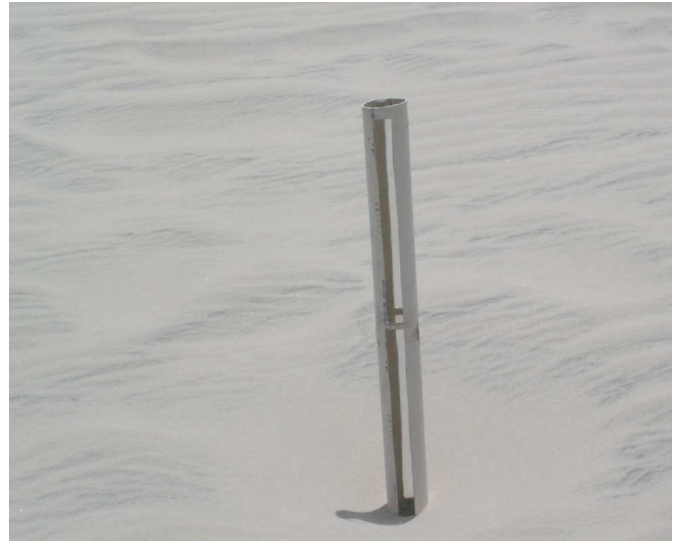}
\caption{Cylindrical sand trap in use on a field dune.}
\label{fig_1}
\end{center}
\end{figure}

Among the many different trap designs that have been used in
the past one of the simplest and most popular is a hollow
cylinder with two slits, one open and the other covered by a
membrane that is impermeable to the grains~\cite{Maia} see
Fig.~\ref{fig_1}. It has been extensively implemented in
field studies along the Northern coast of Brazil with much
success. We will therefore study this particular device in
detail by numerically solving the turbulent flow around an
object having the corresponding geometry. We will calculate
the trajectories of grains released at different positions
in the area in front of the trap to assess if they are
captured or not and do this for various Reynolds and Stokes
numbers as well as different membrane permeabilities.
This paper is organized as follows. In Section $2$, we
introduce the model and briefly summarize the algorithm used
in our simulation. In Section $3$ we present the qualitative
and quantitative results. We conclude in Section $4$ by
discussing the importance of these results in some practical
applications.
\section{Model Formulation}
\begin{figure}
\begin{center}
\includegraphics[width=10.0cm,height=14.0cm]{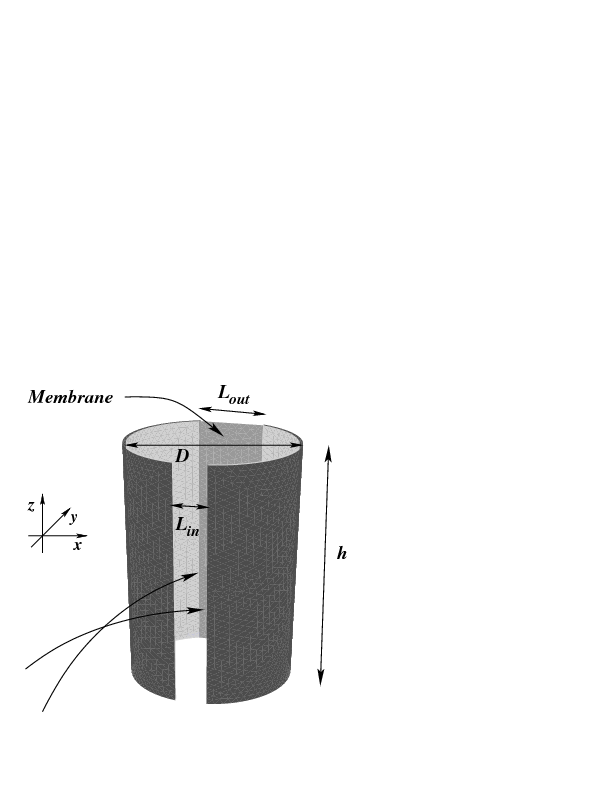}
\caption{The geometry of the sand trap investigated here. $D$ is the 
diameter and $h$ the height. The widths of the vertical
slits at the front and back sides are $L_{in}$ and
$L_{out}$, respectively. Two possible grain trajectories are
sketched.}
\label{fig_2}
\end{center}
\end{figure}
As shown in Fig.~\ref{fig_2} the sand trap that we study
consists of a hollow cylinder of diameter $D$ and height $h$
closed on the top and the bottom, and having two vertical
slits, one at the front and another at the back side. For
simplicity we will neglect the width of the cylinder
walls. A membrane covers the slit of the back side in order
to capture particles with diameters larger than
$0.01~mm$. The slits at the front and back sides have widths
$L_{in}$ and $L_{out}$, respectively. We chose a height $h =
1m$, a diameter $D = 0.05m$ and slit widths $L_{in}=0.01m$
on the front side and $L_{out}=0.02m$ on the back side in
agreement with typical dimensions of real devices. The
membrane has a thickness of $\Delta m=0.001m$. Its
permeability $\alpha$ can be modified and will be one of the
control parameters of our simulation. In order to simulate
the conditions of a realistic wind field we consider a box
of size $1.5 \times 1.5 \times 1 m$ with moving boundaries
on top, left and right, i.e. in all directions transverse to
the wind direction except for the bottom. On the fixed
walls, i.e. the trap surface and the bottom we impose
non-slip boundary conditions, which means that the velocity
is equal to zero at the interfaces between solid and
fluid. In order to reduce the size effects the simulation box
was chosen to be one order of magnitude larger than diameter
$D$ of the trap.

We assume that atmospheric air is an incompressible and
Newtonian fluid having a viscosity of
$\mu=1.7895\times10^{-5}kg~m^{-1}~s^{-1}$ and a density of
$\rho=1.225~kg/m^{3}$. The Reynolds number is defined as
$Re\equiv \rho V L_{x}/\mu$, where $V$ is the average
velocity and $L_{x}$ the linear size of the box in $x$
direction. This gives in our case $Re\approx 40000$
meaning that we can assume to be in a fully developed
homogeneous turbulent state. The standard $k-\epsilon$ model
is therefore an adequate way to solve the corresponding
Reynolds-averaged Navier-Stokes equations of motion. Their
numerical solution can be achieved by discretizing the
velocity and pressure fields and using a control volume
finite-difference technique~\cite{Patankar,Fluent}. The
convergence criteria of this scheme can be defined in terms
of residuals, which measure up to which degree the
conservation laws are satisfied. In our simulation we
consider that convergence is achieved when each of the
normalized residuals is smaller than $10^{-6}$.

A fully turbulent atmospheric boundary layer over a flat
surface shows a logarithmic increase of the velocity $v(z)$
with the distance $z$ from the surface. Therefore, we impose
a logarithmic velocity profile on the inlet, i.e. in the $x
= 0$ plane, of the form
\begin{equation} 
v(z)=\frac{u_{*}}{\kappa}\ln\frac{z}{z_{0}},
\label{eq_1}
\end{equation}
where $u_{*}=0.36 m/s$ denotes the shear velocity,
$z_{0}=1.0 \times 10^{-3}m$ the so called roughness length
and $\kappa=0.4$ is the von K\'arm\'an
constant~\cite{Sauermann}.

We model the fabric covering the opening at the backside of
the trap as a special type of boundary condition mimicking a
porous medium where the velocity/pressure drop characteristics are
known. If $\Delta m$ is the thickness of this membrane
then the pressure drop is defined according to Darcy's law
as:
\begin{equation} 
\Delta p = - \frac{\mu }{\alpha} v \Delta m ,
\label{eq_2}
\end{equation} 
where $\alpha$ is the permeability of the medium and $v$ is
the velocity normal to the membrane. In order to understand
the effect of the membrane on the wind velocity field, we
perform simulations using different membrane permeabilities, namely
$\alpha=1.0 \times 10^{-6}$, $1.0\times 10^{-8}$, $1.0
\times 10^{-9}$ and $1.0 \times10^{-10}$.
\begin{figure*}
\begin{center}
\includegraphics[width=14.0cm,height=18.0cm]{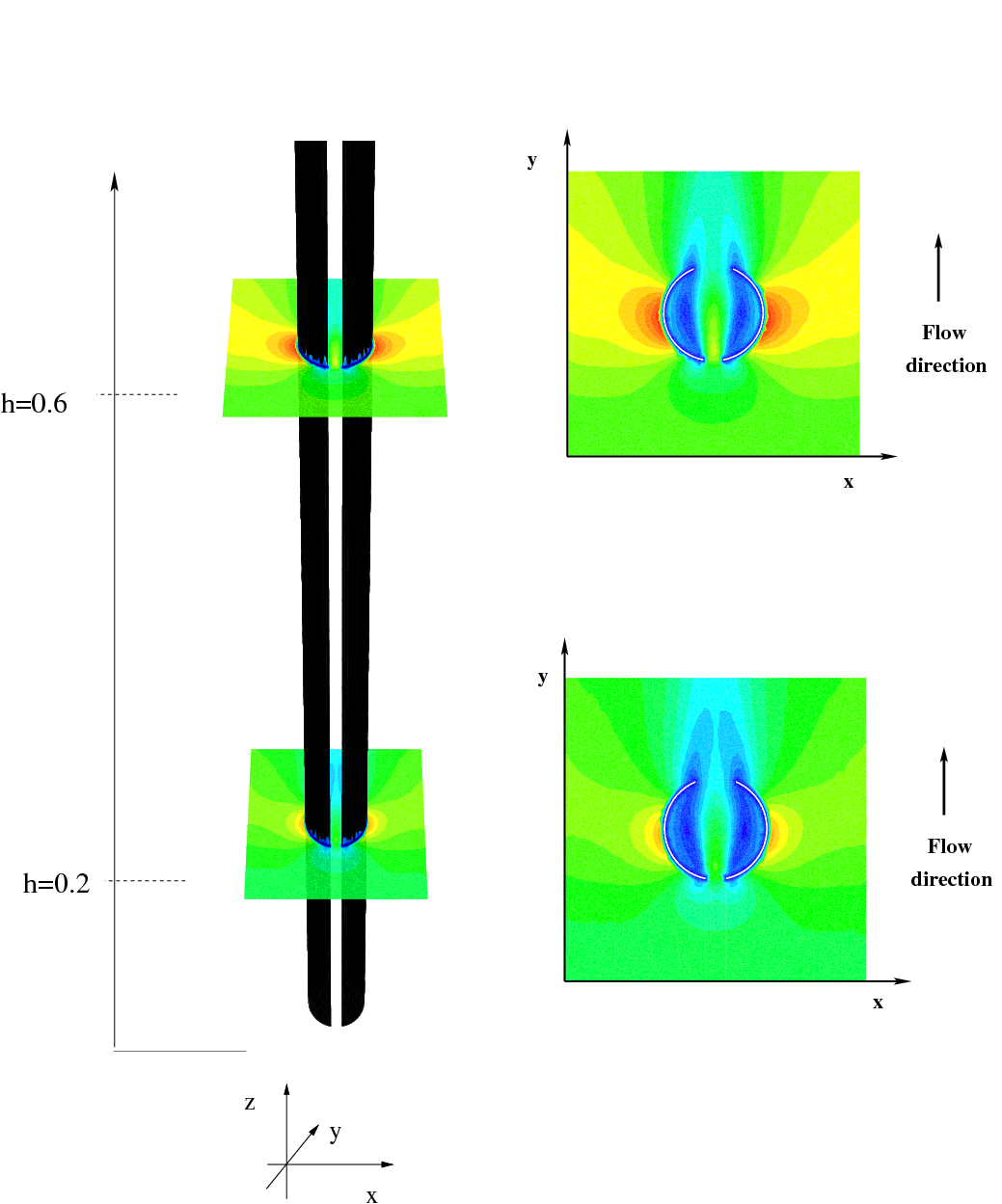}
\caption{The pictures show the velocity profile in and around 
the sand trap. The images at right, show two different
velocity profiles in a two dimensional cut (plane)
through the sand trap at height $z=0.2$ and $z=0.6$. The
wind velocity profile input at the surface $y=0$ comes
form Eq.~(\ref{eq_1}). The colors indicate the magnitude of
the velocity varying from blue (low value) to red (high
value). The picture corresponds to permeabilities
$\alpha=1.0\times 10^{-6}$.}
\label{fig_3}
\end{center}
\end{figure*}

For simplicity we will only consider spherical particles. We
also assume that the density of flying particles is so
low that we can neglect collisions between them. For the
same reason we will also neglect the momentum loss exerted
by the particles on the fluid. Consequently the wind
velocity and pressure fields can be calculated without
knowing the particle positions and one can then obtain the
trajectory of each particle by just integrating its 
equation of motion:
\begin{equation}
m_{p}\frac{d{\bf u}_{p}}{dt}=\sum {\bf F}_{p},
\label{eq_3}
\end{equation}
where $m_{p}$ and ${\bf u}_{p}$ are the mass and the
velocity of a particle of diameter $d_{p}$ and $\sum {\bf
F_{p}}$ is the total force acting on this particle. Let us
assume that drag and gravity are the only relevant forces
and that the particles do not interact with each other. Then
the equation of motion for one particle can be rewritten as
\begin{equation}
\frac{d {\bf u}_{p}}{dt} = F_D({\bf u} - {\bf u}_{p}) + {\bf g} \frac{(\rho_p - \rho)}{\rho_p}, 
\label{eq_4}
\end{equation}	
where ${\bf g}$ is gravity acceleration and
$\rho_{p}=2650~kg~m^{-3}$ is a typical value for the density
of sand particles. The term $F_D({\bf u} - {\bf u}_{p})$ in
Eq.~(\ref{eq_4}) represents the drag force per unit
particle-mass where
\begin{equation} 
F_D = \frac{18 \mu}{\rho_p d^{2}_{p}} \;\; \frac{C_D {\rm Re_{p}}}{24}.
\label{eq_5}
\end{equation}
Here $Re_{p}\equiv \rho d_{p}|{\bf u}-{\bf u}_{p}|/\mu$ is
the particle Reynolds number and for the drag coefficient
$C_{D}$ we use an empirical expression taken from
Ref.~\cite{Morsi}. The trajectories of the particles are
calculated by numerically integrating the equation of
motion Eq.~(\ref{eq_4}).

We characterize the effect of inertia on the air borne grains in the
flow field through the dimensionless {\it Stokes} number
$St$ which is defined by $St\equiv \rho_{p}{d_{p}}^2 V/18L_{x}
\mu$. For $St \gg 1$, inertia will dominate and the
particles will move along straight lines not following the
fluid, i.e. they will move in the air ballistically. For $St
\ll 1$ particles behave as tracers and will perfectly follow
the streamlines of the fluid. In our simulation we will
vary the Stokes number by changing the particle diameter
keeping all the other parameters constant. If a particle
hits the outer surface of the sand trap it will be reflected
elastically. If it hits the inner part of the trap it will
stop moving, i.e. is counted as captured.

\section{Results and Discussions}

We first present a study of the influence of the
permeability $\alpha$ of the membrane on the fluid velocity
field without the presence of particles. In order to
understand essential features of the flow field around and
inside the sand trap, we have performed extensive
simulations by analyzing the flow for several values of
permeability $\alpha$. To see the flow field we focus our
attention on horizontal cuts through the trap, at a
different distances $z=0.2$, and $z=0.6$ from the ground.
\begin{figure}
\begin{center}
\includegraphics[width=14.0cm]{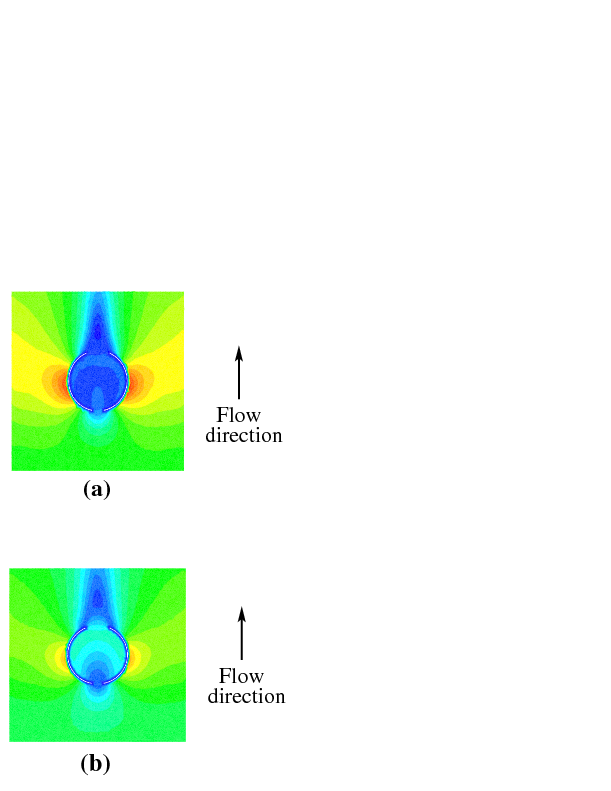}
\caption{The pictures show the velocity profile in and around
the sand trap. The images at right, show two different
velocity profiles in a two dimensional cut through the sand
trap at height $z=0.6$ $(a)$ and $z=0.2$ $(b)$. The wind
velocity profile input comes from Eq.~(\ref{eq_1}). The
colors indicate the magnitude of the velocity varying from
blue (low value) to red (high value). The picture correspond
to permeabilities $\alpha=1.0\times 10^{-10}$.}
\label{fig_4}
\end{center}
\end{figure}
Fig.~\ref{fig_3} and Fig.~\ref{fig_4} depict the velocity
field around the sand trap for two different values of
permeability, $\alpha=1.0 \times 10^{-6}$ and $\alpha=1.0
\times 10^{-10}$, respectively. Here we focus our 
discussion on the velocity profile, in the horizontal cuts
through the trap at different heights $z$. We will first
discuss the case of high permeability. In this case we can
see in Fig.~\ref{fig_3} that the wind velocity presents a
parabolic profile in the channel inside the sand trap and
also the existence of stagnation zones beside this
channel. 
\begin{figure}
\begin{center}
\includegraphics[width=5.0cm]{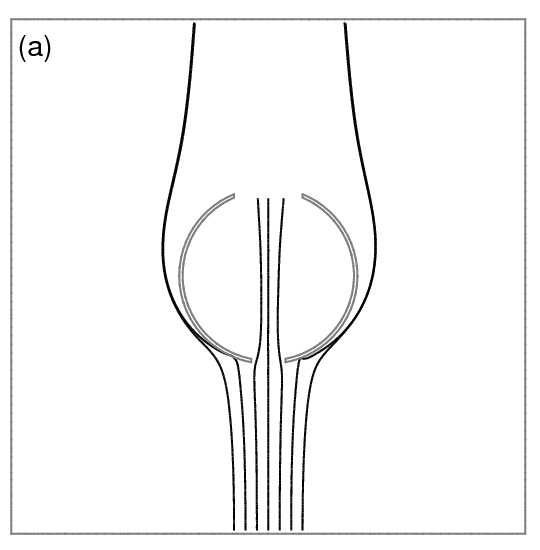}
\includegraphics[width=5.0cm]{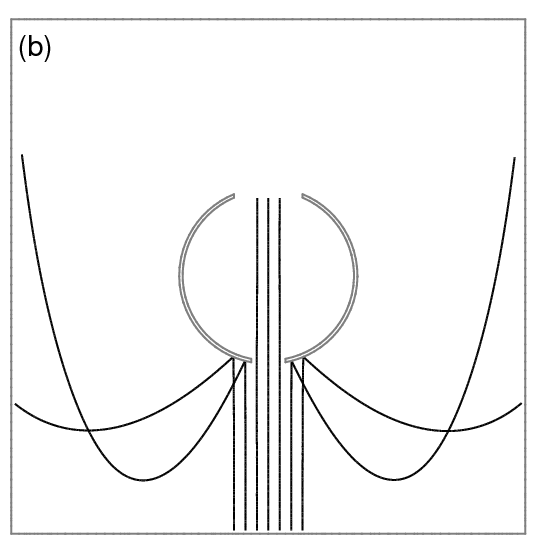}
\includegraphics[width=5.0cm]{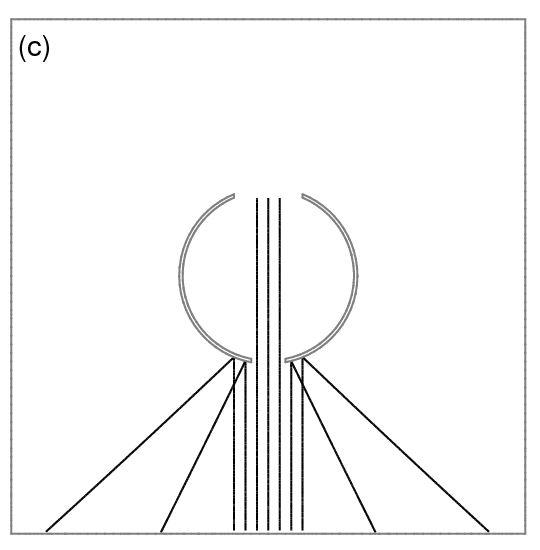}
\caption{The $2D$ view of typical particles trajectories for 
three different Stokes numbers, $(a)$ $St=0.014$,
$(b)$$St=1.43$ and $(c)$ $St=14.34$. The membrane
permeability is $\alpha=1.0 \times 10^{-6}$. The particles
concentrate in the region of high velocities for all values
of $St$.}
\label{fig_5}
\end{center}
\end{figure}
\begin{figure}
\begin{center}
\includegraphics[width=5.0cm]{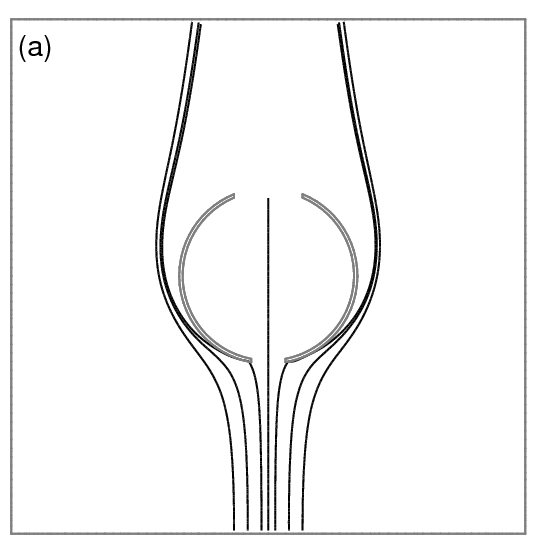}
\includegraphics[width=5.0cm]{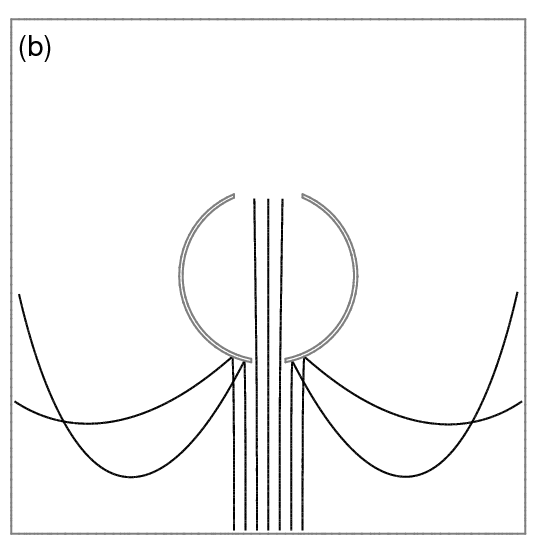}
\includegraphics[width=5.0cm]{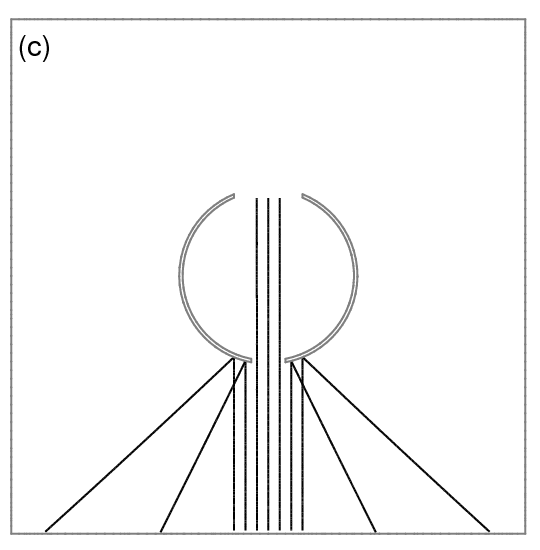}
\caption{Typical particle trajectories projected on the $x-y)$ surface, for 
three different Stokes numbers, $St=0.014 $ $St=1.43$ and
$St=14.34 $ correspond to $(a)$, $(b)$ and $(c)$,
respectively. The value of the membrane permeability is
$\alpha=1.0 \times 10^{-10}$.}
\label{fig_6}
\end{center}
\end{figure}

For low permeability, see Fig.~\ref{fig_4}, the velocity
decreases along the center line to form a zone of low
velocity close to the entrance of the sand trap while it
increases around the trap. This increase of flow is due to
the obstruction caused by the trap to the incompressible
fluid. For larger heights $z$, the zone of low
velocity reaches almost the entire region inside of the sand
trap. With a decreasing membrane permeability also the
formation of shadow zones can be observed behind the sand
trap.

Our central issue is the study of sand transport and in
particular the effect of the membrane permeability on the
sand flux. Therefore we calculate particle trajectories by
extensive simulations for different Stokes numbers and
membrane permeabilities. In order to quantify the capture
process the sand trap efficiency $\eta$ is defined as
\begin{equation}
\eta\equiv\frac{\phi}{\phi_{0}}
\end{equation}
where $\phi_{0}$ is the total number of particles released
during the interval $\Delta x$ and $\phi$ is the number of
particles among these that have been captured. We will
discuss in the following the efficiency $\eta$ as function
of membrane permeability and Stokes number.

\begin{figure}
\begin{center}
\includegraphics*[width=9.0cm]{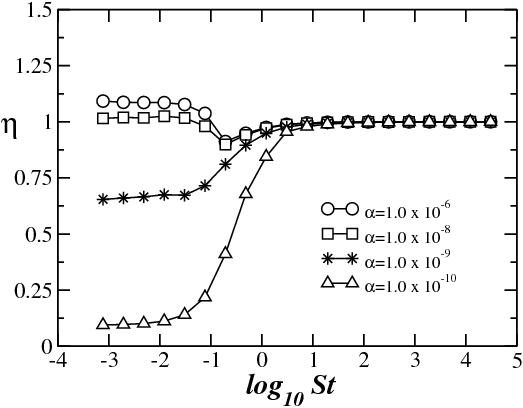}
\caption{Semi-log plot of the efficiency $\eta$ of the 
sand trap as function of the Stokes number. The symbols
correspond to different values of permeabilities $\alpha$,
namely, (circles) $\alpha=1.0\times 10^{-6}$, (squares)
$\alpha=1.0\times 10^{-8}$, (stars) $\alpha=1.0\times
10^{-9}$ and (triangles) $\alpha=1.0\times 10^{-10}$.}
\label{fig_7}
\end{center}
\end{figure}
\begin{figure}
\begin{center}
\includegraphics[width=12.0cm,height=16.0cm]{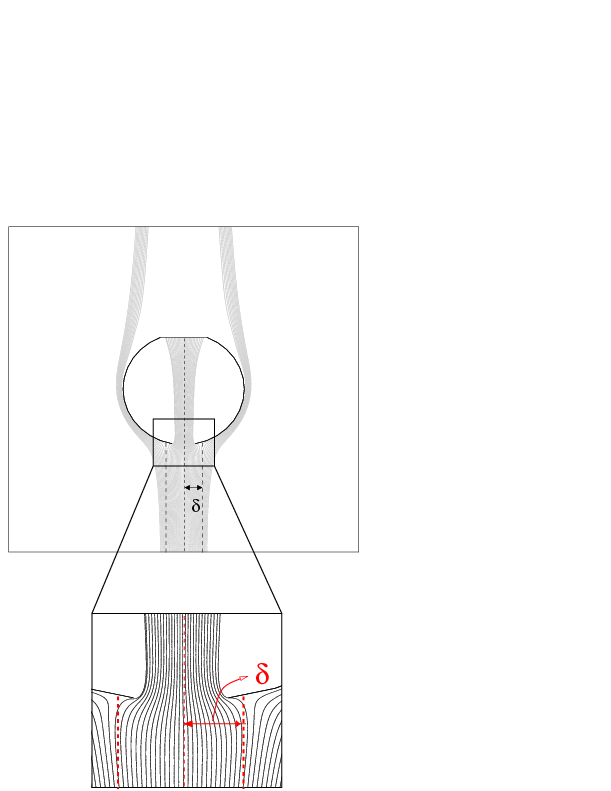}
\caption{2D view of typical particle trajectories 
for high permeability $\alpha=1.0\times 10^{-6}$ and low
Stokes number $St=0.14$ at height $z=0.2$. The distance
$\delta$ is the maximum distance from the
symmetry line for which the particles bend towards the
entrance of the sand trap. The stagnation zones are shown in
more detail in the inset revealing how the particle
trajectories can bend towards the symmetry line of the sand
trap.}
\label{fig_8}
\end{center}
\end{figure}
\begin{figure}
\begin{center}
\includegraphics[width=7.0cm]{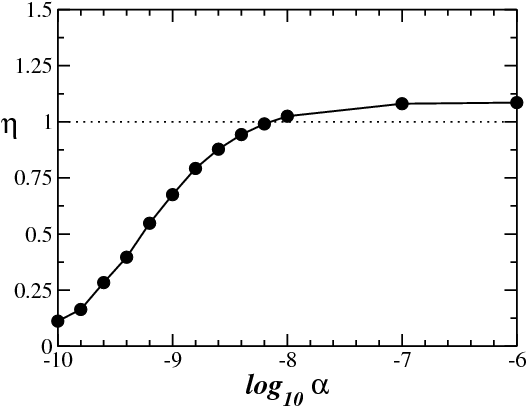}
\caption{Semi-log plot of the efficiency $\eta$ as 
function of the membrane permeability $\alpha$, for fixed
Stokes number $St=1.0\times10^{-2}$. The doted line is included
as guide to eye and corresponds to $\eta=1$. }
\label{fig_9}
\end{center}
\end{figure}

Initially the grains are placed at $y=0$, a given height $z$
and randomly within the interval, $-0.01m< x<0.01m$. For a
fixed value of $St$ and $\alpha$, we release up to $1000$
particles to determine the efficiency of the sand trap. We
show in Figs.~\ref{fig_5} and \ref{fig_6} the trajectories
of $7$ particles for two different permeabilities,
$\alpha=1.0\times 10^{-6}$ and $\alpha=1.0\times 10^{-10}$,
respectively, and different Stokes numbers. For both figures
the Stokes numbers are from top to bottom: $St=0.014$,
$St=1.43$ and $St=14.34$. In the case of high permeability,
as shown in Fig.~\ref{fig_5}, the grains concentrate in the
regions of large velocities. Inside of the trap the particle
trajectories do spread apart more with decreasing Stokes
number. This effect is more pronounced at lower
permeability. For high Stokes numbers some particles collide
with the trap surface revers their direction and leave the
box simulation. For intermediate Stokes numbers these
particles also collide but are then deviated by the wind
flow. This effect is independent of the permeability.

The dependence of the sand trap efficiency $\eta$ on the
Stokes number is shown in Fig.~\ref{fig_7} for different
values of $\alpha$. We performed simulations for membrane
permeabilities $\alpha = 1.0 \times 10^{-6}$, $1.0 \times
10^{-8}$, $1.0 \times 10^{-9}$ and $1.0 \times 10^{-10}$ at
many different heights $z$, with an inlet velocity given by
Eq.~\ref{eq_1}. The value of $\eta$ was obtained from the
average over different heights $z$, where at each one we
released $1000$ particles.  We see that $\eta$ presents two
different regimes as function of the permeability
$\alpha$. Let us first discuss the case of low
permeabilities. In the limit of small Stokes numbers the
efficiency of the sand trap is small and remains essentially
constant close to zero. Since $St\ll 1$ the particles can be
considered as tracers that nearly exactly follow the
streamlines of the flow, avoiding trapping because
practically no streamline crosses the interior of the trap
as consequence of the low membrane permeability. Indeed we
observe in these cases stagnation zones inside the sand
trap. Above $St\approx 0.05$ the efficiency $\eta$
increases as function of the Stokes number. For high values
of $St$ the efficiency reaches a saturation value close to
unity. In this limiting case the particles move
ballistically towards the sand trap and those that have been
released within the range $-0.005m< x<0.005m$ are captured.

For high permeabilities $\alpha$, the efficiency $\eta$
presents an unexpected behavior in the region of low $St$,
as shown in Fig.~\ref{fig_7}. For large Stokes numbers the
efficiency $\eta$ remains constant until $St\approx
10.0$. Below this value surprisingly we detect a small
minimum followed by an increase to a value that can be above
one. We can explain this behavior observing the particle
trajectories in the region close to the entrance of the sand
trap as shown in Fig.~\ref{fig_8}.

For small membrane permeability the sand trap behaves like a
solid cylinder. In this case a stagnation region appears in
front of the trap around the symmetry line. As the
permeability $\alpha$ increases some air can cross the sand
trap and two stagnation zones appear beside the entrance and
outside of the trap. Each of them has a separation point and
all particle trajectories between these two separation
points bend towards the center and enter the trap as
confirmed in the simulation for low Stokes number as shown
in Fig.~\ref{fig_8}. Therefore the number of captured
particles and the efficiency of the sand trap can increase
beyond the ballistic case.
\vspace{2cm}

For a fixed value of $St$ and different permeabilities
$\alpha$, we search the maximum distance from the symmetry
line $\delta$ at which the particles bend towards the
entrance of the sand trap. This $\delta$ can be used as a
measure for the increase in the sand trap efficiency. In
Fig.~\ref{fig_9}, we show in a semi-log plot the efficiency
$\eta$, which has been calculated from the parameter
$\delta$, against the membrane permeability
$\alpha$. Clearly, the curves display a strong change starting at
the permeability $\alpha=1.0\times10^{-8}$. This is the
smallest permeability which still affects the efficiency of
the sand trap.

\section{Conclusions}
In this paper, we have studied numerically the behavior of a
sand trap frequently used to measure aeolian sand transport
in the field. We solved the turbulent wind velocity field in
the presence of the sand trap and investigated the effect of
the membrane permeability. We studied quantitatively the
particle trajectories carried by the fluid for different
membrane permeabilities.

We have shown how the efficiency of the sand trap depends on
the Stokes number and gave some insight about the effect of
the membrane permeability on the capturing process.

As previously observed the membrane permeability strongly
influences the capture process at low Stokes numbers. The
sand trap efficiency $\eta$ exhibits a surprising increase
above one, for large permeabilities and small Stokes
numbers. Let us point out that only at higher $St$ values,
when the trajectories of the particles are completely
ballistic, the efficiency $\eta$ becomes independent on the
membrane permeability. This is a useful result since for
natural sand, for example on dunes, the grain size is around
$200\mu m$ which corresponds to a rather high Stokes
number. Our results therefore confirm that this type of sand
trap is adequate to measure aeolian sand flux. The area of
the slit at the entrance essentially equals the cross
section over which the sand flux is measured independently on
the details of the membrane on the back side.

\section{Acknowledgments}
We would like to thank Andr\'e Moreira for helpful
discussions. We acknowledge CNPq, CAPES, FINEP and FUNCAP/CNPq/PPP
and the Max Planck prize for financial support.


\begin{thebibliography}{99}
\bibitem{Fluent} The FLUENT (trademark of FLUENT Inc.) commercial
package for fluid dynamics analysis is used in this calculation.
\end{thebibliography}
\end{document}